\documentclass[sigconf]{acmart}
\AtBeginDocument{%
  }

\copyrightyear{2026}
\acmYear{2026}
\setcopyright{cc}
\setcctype{by}
\acmConference[CIKM '26]
  {Proceedings of the 35th ACM International Conference on Information and Knowledge Management}
  {November 07--11, 2026}
  {Rome, Italy}
\acmBooktitle{Proceedings of the 35th ACM International Conference on Information and Knowledge Management
  (CIKM '26), November 07--11, 2026, Rome, Italy}
\acmDOI{10.1145/3799682.3839929}
\acmISBN{979-8-4007-2539-5/2026/11}
\begin{document}

\title[Online RL Fine-Tuning for Retrieval]{Learning from What You Retrieve: Online RL Fine-Tuning for Semantic Retrieval}

\author{Shaowei Wei}
\authornote{Corresponding author.}
\orcid{0009-0004-9473-0895}
\email{leqian.wsw@taobao.com}
\affiliation{%
  \institution{Alibaba Group}
  \city{Hangzhou}
  \country{China}
}

\author{Chong Huang}
\orcid{0009-0002-3701-541X}
\email{caiyu.hc@alibaba-inc.com}
\affiliation{%
  \institution{Alibaba Group}
  \city{Hangzhou}
  \country{China}
}

\author{Songtao Fang}
\orcid{0009-0001-7713-0415}
\email{fangsongtao.fst@taobao.com}
\affiliation{%
  \institution{Alibaba Group}
  \city{Hangzhou}
  \country{China}
}

\author{Jin Zhang}
\orcid{0009-0006-4073-7869}
\email{zj146372@alibaba-inc.com}
\affiliation{%
  \institution{Alibaba Group}
  \city{Hangzhou}
  \country{China}
}

\author{Zhuojun Wang}
\orcid{0009-0008-8685-4467}
\email{jerry.wangzj@taobao.com}
\affiliation{%
  \institution{Alibaba Group}
  \city{Hangzhou}
  \country{China}
}

\author{Chengfu Huo}
\orcid{0000-0003-3611-6921}
\email{chengfu.huocf@taobao.com}
\affiliation{%
  \institution{Alibaba Group}
  \city{Hangzhou}
  \country{China}
}

\renewcommand{\shortauthors}{Shaowei Wei et al.}

\begin{abstract}
In large-scale e-commerce retrieval, dual-encoder retrievers are optimized for contrastive similarity, whereas downstream rerankers capture finer-grained relevance preferences; this objective mismatch limits end-to-end retrieval quality. Reinforcement Learning offers a way to use reward-model feedback for retriever adaptation, but we observe that standard policy-gradient updates can degrade embedding geometry, especially when the document index must remain frozen due to industrial constraints.

To address this, we propose \textbf{PAO (Positive-Advantage-Only)}, a selective RL optimization method. Our analysis reveals that indiscriminate penalization of negative samples (pushing away) in a frozen high-dimensional space disrupts pre-trained semantic manifolds. PAO selectively applies gradient updates only to retrieved items with positive advantages, effectively pulling query embeddings toward high-reward regions while preserving global topological stability. Experiments on both a massive industrial dataset and public benchmarks demonstrate that PAO significantly outperforms standard RL and distillation baselines.
\end{abstract}



\maketitle
\raggedbottom
\AddToHookNext{shipout/after}{\flushbottom}

\section{Introduction}

Modern retrieval systems commonly use a two-stage architecture: retrieval followed by ranking \cite{Covington16}. Dual-encoder models \cite{Huang13DSSM} retrieve candidates efficiently with approximate nearest-neighbor search \cite{Johnson19}, but often miss the complex, fine-grained relevance signals modeled by heavy downstream rerankers \cite{Burges05, Nogueira19}.

Bridging this gap via online fine-tuning is challenging in industrial settings, where optimization must operate with a \textbf{frozen document index}. Re-indexing billions of item vectors after every model update is computationally prohibitive, so optimization is constrained to the query encoder alone. We formulate this alignment task as a Reinforcement Learning \cite{Sutton18} problem, where the query encoder acts as an agent and receives reward signals from the downstream reranker \cite{Ai18, Chen19}.

However, we identify a critical failure mode: standard Policy Gradient (PG) algorithms lead to severe performance deterioration, a phenomenon we term as \textbf{Geometry Collapse}. Our investigation suggests that the point-wise optimization of RL---specifically pushing away queries from low-reward items---acts destructively when the document space is immutable. Unlike contrastive learning where negatives are pushed relative to positives \cite{Chen21, He20}, pure RL pushing in a frozen space often displaces query embeddings into semantically ambiguous regions, shattering the distribution established by pre-training \cite{Oord18, Xiao23}.

To resolve this, we propose \textbf{PAO}, a selective update mechanism for applying positive-advantage-only RL under the frozen-index constraint. We hypothesize that strictly pulling queries toward candidates with \textit{positive advantages} (rewards above the baseline) allows for fine-grained local alignment, while masking negative signals prevents structural disruption.

\textbf{Our contributions are:}
\begin{enumerate}
    \item We identify the geometric degradation caused by standard RL when fine-tuning query encoders against a frozen index.
    \item We propose PAO, a selective update mechanism that harmonizes retrieval-ranking consistency without compromising embedding space generalization.
    \item Extensive experiments on industrial and public datasets confirm PAO's superiority in both standard metrics and semantic alignment.
\end{enumerate}

\section{Related Work}
\begin{figure*}[!t]
  \centering
  \includegraphics[width=\textwidth]{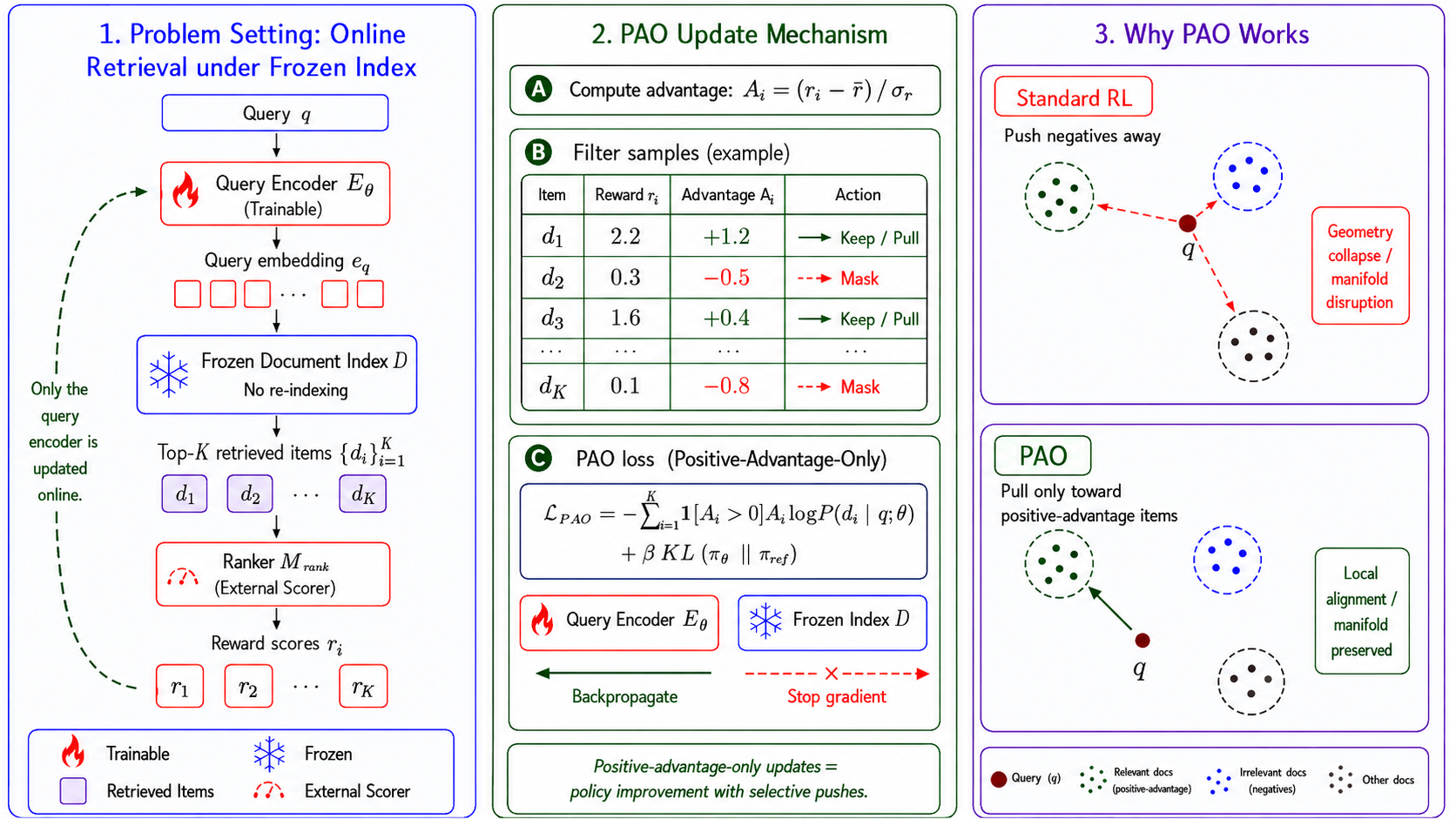}
  \caption{The Query Encoder interacts with the frozen Document Index. Only gradients from positive advantage samples are back-propagated, while negative signals are masked to prevent manifold disruption.}
  \Description{This figure illustrates the architecture of the selective reinforcement learning framework for semantic retrieval. It shows the Query Encoder as a policy network interacting with an online retrieval module, receiving rewards from a reward model. A key component is the selective optimization, where only positive advantage signals are used for gradient updates. A KL divergence loss is also included to maintain embedding space stability.}
  \label{fig:overall_framework}
\end{figure*}

Dense retrieval has been advanced by pre-trained language models and contrastive objectives \cite{Lin21,Oord18,Xiao23}, with representative systems improving negative mining, embedding structure, and retrieval efficiency \cite{Karpukhin20,Xiong20,Li23GTE,Zhang24}. However, most methods assume that both query and document encoders can be updated, or that document embeddings can be rebuilt after training. This assumption is often infeasible in industrial retrieval, where the document index is frozen and only the query encoder can be adapted online.

Retriever-reranker alignment is commonly approached through knowledge distillation, where cross-encoder or reranker scores supervise a dual encoder \cite{Zeng20,Qu21,Ren21,Chen24BGEM3}. Such methods typically distill static teacher distributions over fixed candidate sets, whereas our setting uses feedback on the candidates retrieved by the current query policy. RL has also been studied for search and ranking \cite{Ai18,Chen19,Singh18,Oosterhuis18}, and recent work explores reinforcement learning for dense retrieval or reranker optimization \cite{Tao25,Zhuang26RankR1,Xu25,Lin25}. In contrast, we focus on query-only RL under a frozen document index, where negative policy-gradient updates can distort the embedding geometry.

\section{Methods}

We formulate the fine-tuning of the Query Encoder $E_Q(\cdot ;\theta)$ against a frozen Document Index $\mathcal{D}$ and an external reward model $M_{\mathrm{rank}}$ as an episodic Markov Decision Process (MDP). The proposed framework is depicted in Figure~\ref{fig:overall_framework}.

\subsection{RL Formulation}
\textbf{Policy ($\pi_\theta$):} Defined by the Query Encoder. Given query $q$, it generates $e_q = E_Q(q; \theta)$.\\
\textbf{Action ($a$):} The retrieval of a Top-K list $L_q = \{d_1, \dots, d_k\}$ from $\mathcal{D}$ via MIPS. This list is treated as a macro-action sampled from the policy.\\
\textbf{Sampling Probability:} The probability of sampling document $d_i$ is proportional to the similarity score $s_i = e_q^\top e_{d_i}$. We model the intra-list distribution via Softmax:
\begin{equation}
  P(d_i | q; \theta) = \frac{\exp(s_i / \tau)}{\sum_{j \in L_q} \exp(s_j / \tau)}
  \label{eq:1}
\end{equation}
where $\tau$ is the temperature coefficient.

\subsection{Reward and The Geometry Collapse}
The ranker supplies the relevance reward $r_i=M_{\mathrm{rank}}(q,d_i)$. We utilize standardized advantage $A_i = (r_i - \bar{r})/\sigma_r$ to reduce variance \cite{Shao24}.
The standard REINFORCE objective \cite{Williams92} minimizes:
\begin{equation}
  \mathcal{L}_{all} = - \sum_{i=1}^k A_i \log P(d_i | q; \theta)
  \label{eq:2}
\end{equation}

\textbf{Observation:} When $A_i < 0$, the gradient drives $e_q$ away from $e_{d_i}$. In a frozen high-dimensional space, this pushing operation has infinite degrees of freedom. Unlike contrastive learning (which pulls negatives towards other defined clusters), pure RL pushing often displaces $e_q$ into semantic voids, disrupting the cluster structure required for recall.


\subsection{Positive-Advantage-Only (PAO) Strategy}
To incorporate ranking knowledge without destroying semantic structure, we propose PAO. We mask negative updates and enforce a KL-divergence constraint relative to the pre-trained reference policy $\pi_{ref}$:

\begin{equation}
  \mathcal{L}_{pos}(\theta) = - \sum_{i=1}^k \mathbb{I}(A_i > 0) \cdot A_i \cdot \log P(d_i | q; \theta) + \beta \cdot \text{KL}(\pi_\theta || \pi_{ref})
  \label{eq:3}
\end{equation}

Here, $\mathbb{I}(\cdot)$ is the indicator function. By optimizing only when $A_i > 0$, we effectively pull the query embedding towards items that the reward model considers \textit{better than average}. This constitutes a constructive local adjustment, whereas the KL term \cite{Ouyang22} (weighted by $\beta$) acts as a global anchor to prevent policy drift.

\definecolor{darkgreen}{RGB}{0,128,0}
\begin{table*}[h]
  \centering
  \caption{Retrieval Performance on Industrial Dataset}
  \label{tab:industrial_results}
  \begin{tabular}{ccccccccc}
    \toprule
    \textbf{Model} & \textbf{Recall@5} & \textbf{Recall@10} & \textbf{Recall@20} & \textbf{Recall@50} & \textbf{NDCG@5} & \textbf{NDCG@10} & \textbf{NDCG@20} & \textbf{NDCG@50} \\[0.5ex]
    \midrule
    Baseline & 0.7929 & 0.8530 & 0.9002 & 0.9431 & 0.7000 & 0.7196 & 0.7316 & 0.7402 \\[0.5ex]
    RL-All & 0.6547 & 0.7222 & 0.7825 & 0.8496 & 0.5635 & 0.5854 & 0.6006 & 0.6141 \\
           & \textcolor{darkgreen}{(-13.8pt)} & \textcolor{darkgreen}{(-13.1pt)} & \textcolor{darkgreen}{(-11.8pt)} & \textcolor{darkgreen}{(-9.4pt)} & \textcolor{darkgreen}{(-13.6pt)} & \textcolor{darkgreen}{(-13.4pt)} & \textcolor{darkgreen}{(-13.1pt)} & \textcolor{darkgreen}{(-12.6pt)} \\[0.5ex]
    \textbf{RL-Pos (PAO)} & \textbf{0.8618} & \textbf{0.9030} & \textbf{0.9341} & \textbf{0.9623} & \textbf{0.7896} & \textbf{0.8030} & \textbf{0.8109} & \textbf{0.8165} \\
                & \textcolor{red}{(+6.9pt)} & \textcolor{red}{(+5.0pt)} & \textcolor{red}{(+3.4pt)} & \textcolor{red}{(+1.9pt)} & \textcolor{red}{(+9.0pt)} & \textcolor{red}{(+8.3pt)} & \textcolor{red}{(+7.9pt)} & \textcolor{red}{(+7.6pt)} \\[0.5ex]
    \bottomrule
  \end{tabular}
\end{table*}

\section{Experiments}

\subsection{Experimental Setup}
\label{sec:experimental_setup}

\textbf{Dataset:} We evaluate our method on a proprietary, anonymized search log from a top-tier e-commerce platform, consisting of 1M training queries and 50k test queries.\\
\textbf{Model Configurations:} The query and document encoders are initialized with GTE-Base \cite{Li23GTE,Zhang24}. For reinforcement learning, we employ an online fine-tuned BGE-Reranker \cite{Chen24BGEM3} as the reward model.\\
\textbf{Training Settings:} For PAO, we freeze the document index, update only the query encoder, and retrieve top-100 candidates per query. We train for two epochs with batch size 16, learning rate $2\times10^{-5}$, 100 warmup steps, and $\beta=0.3$.\\
\textbf{Compared Methods:} We compare the following strategies:
\begin{itemize}
    \item \textbf{Baseline:} The GTE-Base model fine-tuned using standard contrastive learning (InfoNCE) on in-domain query-document pairs.
    \item \textbf{RL-All:} A reinforcement learning baseline using the standard Listwise Policy Gradient on all document samples.
    \item \textbf{RL-Pos (PAO):} The proposed selective RL method that applies updates only from positive-advantage samples.
\end{itemize}

\subsection{Main Results: Industrial Dataset}
We evaluate all methods on standard information retrieval metrics. Table \ref{tab:industrial_results} presents the results on our industrial dataset.


Consistent with our hypothesis, \textbf{RL-All} suffers a catastrophic drop (e.g., -13.6pt in NDCG@5). This confirms that unselective optimization harms representation quality. When negative gradients push the query vector, they do so blindly without knowledge of the manifold structure, often pushing the query into a region of the vector space that corresponds to no meaningful semantic concept.

In contrast, \textbf{RL-Pos (PAO)} achieves robust improvements (+9.0pt in NDCG@5 over Baseline). By relying solely on pull operations, PAO ensures that the query vector is always moving towards a valid document vector (a known point in the semantic manifold), thereby implicitly preserving the validity of the query representation.

\subsection{Semantic Consistency Analysis via LLM-Judge}
We further use Qwen3-235b-a22b \cite{Yang25} as an LLM judge to assess whether retrieved items fully satisfy query intent. Hits@K counts relevant items within the top-K results, while Matchment measures average demand-point satisfaction across the retrieved list.

\begin{table}[!h]
  \centering
  \caption{Objective Retrieval Quality Evaluation Under Different Fine-Tuning Approaches}
  \label{tab:llm_judge_results}
  \resizebox{\columnwidth}{!}{%
  \begin{tabular}{ccccc}
    \toprule
    \textbf{Model} & \textbf{Hits@5} & \textbf{Hits@10} & \textbf{Hits@20} & \textbf{Matchment} \\[0.5ex]
    \midrule
    Baseline & 0.3289 & 0.2631 & 0.2130 & 0.6764 \\[0.5ex]
    RL-All & 0.2822 & 0.2306 & 0.1901 & 0.6558 \\
           & \textcolor{darkgreen}{(-4.7pt)} & \textcolor{darkgreen}{(-3.2pt)} & \textcolor{darkgreen}{(-2.3pt)} & \textcolor{darkgreen}{(-2.1pt)} \\[0.5ex]
    \textbf{RL-Pos (PAO)} & \textbf{0.3458} & \textbf{0.2831} & \textbf{0.2337} & \textbf{0.6900} \\
                & \textcolor{red}{(+1.7pt)} & \textcolor{red}{(+2.0pt)} & \textcolor{red}{(+2.1pt)} & \textcolor{red}{(+1.4pt)} \\[0.5ex]
    \bottomrule
  \end{tabular}
  }
\end{table}

As shown in Table \ref{tab:llm_judge_results}, RL-All degrades across all judged metrics, confirming that unconstrained negative updates blur fine-grained intent matching. By contrast, RL-Pos (PAO) improves deep retrieval (Hits@20 +2.1pt) and Matchment (+1.4pt), suggesting that PAO retrieves harder relevant items without drifting into irrelevant regions.

\subsection{Visual Analysis of Geometry}
\begin{figure*}[!t]
  \centering
  \includegraphics[width=\textwidth]{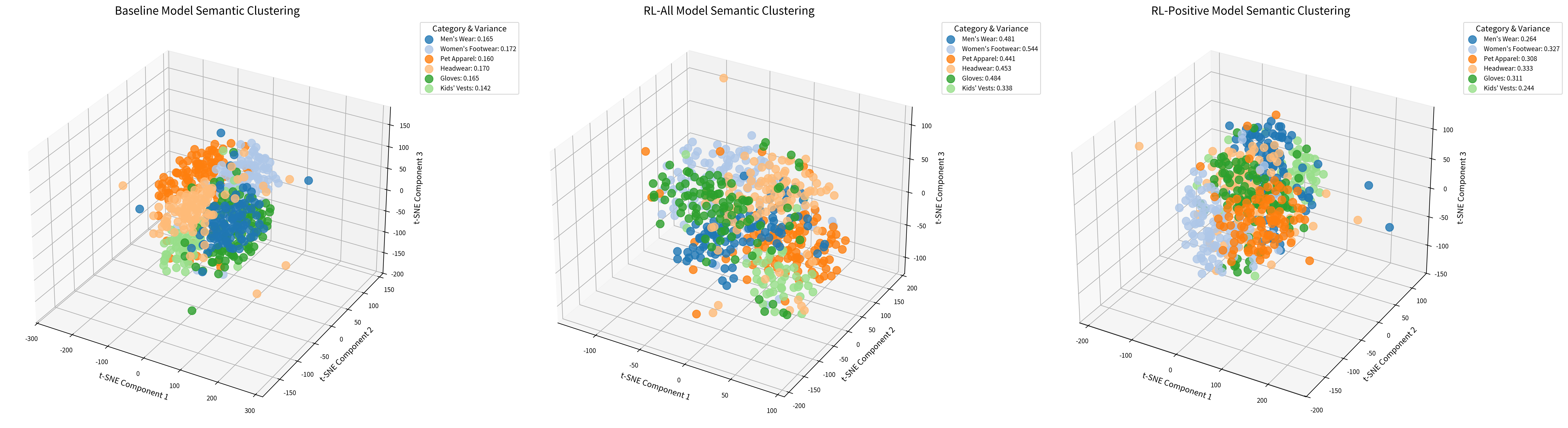}
  \caption{Left: Baseline clusters. Middle: RL-All showing dispersed/collapsed clusters. Right: RL-Pos (PAO) retaining macro-structure with beneficial intra-cluster variance.}
  \Description{This figure displays t-SNE visualizations of query embeddings. The left panel shows the Baseline model with well-separated clusters for different categories. The middle panel, representing the RL-All model, shows highly dispersed and overlapping clusters, indicating a degraded semantic structure. The right panel, for the RL-Pos (PAO) model, largely retains the clustered structure of the Baseline but with slight internal adjustments, suggesting improved fine-grained alignment.}
  \label{fig:tsne_visualization}
\end{figure*}

To visualize the impact of fine-tuning on Query Embedding space geometry, we conducted a t-SNE \cite{Maaten08} analysis. We randomly sampled hundreds of queries from 6 item categories and projected Query Embeddings from Baseline, RL-All, and RL-Pos (PAO) models into 3D space, as shown in Figure \ref{fig:tsne_visualization}.

The Baseline exhibits a robust semantic structure with compact query clusters, a direct result of contrastive pre-training. In stark contrast, RL-All displays significant representation degradation, characterized by dispersed category clusters and a sharp increase in intra-class variance. This empirically confirms that the unconstrained penalization of negative samples destructively distorts the embedding space. Crucially, RL-Pos (PAO) largely preserves the Baseline's macro-structure, demonstrating a non-destructive optimization process. We interpret the observed slight increase in intra-cluster variance not as degradation, but as beneficial adaptation: query embeddings are finely shifted from general category centroids toward high-reward sub-regions within the frozen document space, yielding more discriminative representations.

\subsection{Generalization on Public Benchmarks}
To verify the robustness of PAO beyond proprietary data, we conducted experiments on \textbf{MS MARCO} passage ranking \cite{Bajaj16}, a large-scale public retrieval benchmark. We initialize the retriever from GTE-Base \cite{Li23GTE,Zhang24} and use BGE-Reranker \cite{Chen24BGEM3} as the reward model, using the same frozen-index, query-only setup and training settings as in the industrial experiments described in Section~\ref{sec:experimental_setup}.


As shown in Table \ref{tab:msmarco_results}, PAO improves all reported metrics on MS MARCO, including +2.03pt on NDCG@5 and +4.23pt on Recall@50 over the baseline retriever. This indicates that PAO transfers beyond the proprietary setting and remains effective when aligning a general-purpose retriever with a strong reranker reward model on a large public benchmark.

\begin{table*}[!t]
  \centering
  \caption{MS MARCO Results with Baseline and Distillation Comparisons}
  \label{tab:msmarco_results}
  \begin{tabular}{lcccccccc}
    \toprule
    \textbf{Model} & \textbf{Recall@5} & \textbf{Recall@10} & \textbf{Recall@20} & \textbf{Recall@50} & \textbf{NDCG@5} & \textbf{NDCG@10} & \textbf{NDCG@20} & \textbf{NDCG@50} \\
    \midrule
    Baseline & 0.3579 & 0.4572 & 0.5683 & 0.6922 & 0.2533 & 0.2860 & 0.3145 & 0.3396 \\
    KL-Distill & 0.3803 & 0.4945 & 0.5951 & 0.7114 & 0.2727 & 0.3102 & 0.3360 & 0.3595 \\
    \textbf{RL-Pos (PAO)} & \textbf{0.3850} & \textbf{0.5040} & \textbf{0.6100} & \textbf{0.7345} & \textbf{0.2736} & \textbf{0.3125} & \textbf{0.3398} & \textbf{0.3651} \\
    $\Delta$ vs. Baseline & \textcolor{red}{(+2.71pt)} & \textcolor{red}{(+4.68pt)} & \textcolor{red}{(+4.17pt)} & \textcolor{red}{(+4.23pt)} & \textcolor{red}{(+2.03pt)} & \textcolor{red}{(+2.65pt)} & \textcolor{red}{(+2.53pt)} & \textcolor{red}{(+2.55pt)} \\
    $\Delta$ vs. KL-Distill & \textcolor{red}{(+0.47pt)} & \textcolor{red}{(+0.95pt)} & \textcolor{red}{(+1.49pt)} & \textcolor{red}{(+2.31pt)} & \textcolor{red}{(+0.09pt)} & \textcolor{red}{(+0.23pt)} & \textcolor{red}{(+0.38pt)} & \textcolor{red}{(+0.56pt)} \\
    \bottomrule
  \end{tabular}
\end{table*}

\subsection{Ablation Study}
We investigate two complementary factors: whether PAO is preferable to directly distilling reranker scores on the public benchmark, and how the KL penalty weight ($\beta$) and temperature ($\tau$) affect the industrial setting.

\textbf{Comparison with KL Distillation:} KL-Distill uses the same top-100 candidates, BGE-Reranker teacher, and training settings, but matches the retriever distribution to the reranker-induced soft distribution. As Table~\ref{tab:msmarco_results} shows, PAO remains stronger, especially at deeper cutoffs (+2.31pt Recall@50 and +0.56pt NDCG@50), indicating that positive-advantage filtering converts reranker feedback more stably than direct KL matching under a frozen index.

\textbf{Impact of KL Weight ($\beta$):} Table~\ref{tab:ablation_beta} shows that a moderate $\beta=0.3$ is optimal. Lower values lead to overfitting ranking noise, while higher values ($\beta=1.0$) overly constrain the policy.

\textbf{Impact of Temperature ($\tau$):} Table~\ref{tab:ablation_tau} indicates $\tau=1.0$ provides the best trade-off. Lower $\tau$ reduces exploration, while higher $\tau$ dilutes the gradient signal.

\begin{table*}[!t]
  \begin{minipage}[t]{0.49\textwidth}
    \centering
    \caption{Sensitivity to KL Weight ($\beta$)}
    \label{tab:ablation_beta}
    \begin{tabular}{ccccccc}
      \toprule
      $\beta$ & 0 & 0.1 & 0.3 & 0.5 & 0.7 & 1.0 \\
      \midrule
      Recall@5 & 0.813 & 0.834 & \textbf{0.862} & 0.820 & 0.810 & 0.803 \\
      \bottomrule
    \end{tabular}
  \end{minipage}\hfill
  \begin{minipage}[t]{0.46\textwidth}
    \centering
    \caption{Sensitivity to Temperature ($\tau$)}
    \label{tab:ablation_tau}
    \begin{tabular}{cccccc}
      \toprule
      $\tau$ & 0.3 & 0.5 & 1.0 & 1.5 & 2.0 \\
      \midrule
      Recall@5 & 0.854 & 0.850 & \textbf{0.862} & 0.858 & 0.843 \\
      \bottomrule
    \end{tabular}
  \end{minipage}
\end{table*}

\subsection{Discussion \& Limitations}
While PAO successfully mitigates embedding space degradation, it operates under the assumption that the frozen index covers the recall target. The method optimizes retrievability but cannot generate new knowledge; if relevant documents are absent from the index, PAO serves limited utility. Additionally, our framework assumes the Reward Model is a reliable proxy for user satisfaction. In scenarios where the ranker exhibits bias (e.g., favoring click-bait), PAO may efficiently propagate these biases into the retrieval stage. Addressing this reward hacking via multi-objective optimization remains a direction for future work.

\section{Conclusion}

This paper presents a novel online fine-tuning framework for retrieval models under the strict constraint of a frozen document index. We demonstrate that the \textbf{Positive-Advantage-Only} strategy effectively prevents the geometry collapse associated with standard RL by avoiding destructive updates from negative samples. Beyond immediate performance gains, PAO suggests a potential path toward closed-loop retriever-ranker adaptation: query-side retriever updates may expose harder and more relevant candidates to the ranker, while future ranker updates may provide sharper reward signals for subsequent retriever fine-tuning. We leave a full validation of this iterative loop to future work.

\section*{GenAI Usage Disclosure}
The authors used generative AI tools to assist with language polishing, manuscript organization, visualization prototyping, and code debugging. The authors reviewed and validated all technical claims, references, figures, and experimental results. Generative AI was not used to fabricate data or alter evaluation outcomes.

\bibliographystyle{ACM-Reference-Format}

\begin{thebibliography}{33}


\ifx \showCODEN    \undefined \def \showCODEN     #1{\unskip}     \fi
\ifx \showISBNx    \undefined \def \showISBNx     #1{\unskip}     \fi
\ifx \showISBNxiii \undefined \def \showISBNxiii  #1{\unskip}     \fi
\ifx \showISSN     \undefined \def \showISSN      #1{\unskip}     \fi
\ifx \showLCCN     \undefined \def \showLCCN      #1{\unskip}     \fi
\ifx \shownote     \undefined \def \shownote      #1{#1}          \fi
\ifx \showarticletitle \undefined \def \showarticletitle #1{#1}   \fi
\ifx \showURL      \undefined \def \showURL       {\relax}        \fi
\providecommand\bibfield[2]{#2}
\providecommand\bibinfo[2]{#2}
\providecommand\natexlab[1]{#1}
\providecommand\showeprint[2][]{arXiv:#2}

\bibitem[Ai et~al\mbox{.}(2018)]%
        {Ai18}
\bibfield{author}{\bibinfo{person}{Qingyao Ai}, \bibinfo{person}{Keping Bi},
  \bibinfo{person}{Jiafeng Guo}, {and} \bibinfo{person}{W.~Bruce Croft}.}
  \bibinfo{year}{2018}\natexlab{}.
\newblock \showarticletitle{Learning a deep listwise context model for ranking
  refinement}. In \bibinfo{booktitle}{\emph{Proceedings of the 41st
  International {ACM} {SIGIR} Conference on Research and Development in
  Information Retrieval}}. \bibinfo{publisher}{{ACM}},
  \bibinfo{pages}{135--144}.
\newblock
\href{https://doi.org/10.1145/3209978.3209985}{doi:\nolinkurl{10.1145/3209978.3209985}}


\bibitem[Bajaj et~al\mbox{.}(2016)]%
        {Bajaj16}
\bibfield{author}{\bibinfo{person}{Payal Bajaj}, \bibinfo{person}{Daniel
  Campos}, \bibinfo{person}{Nick Craswell}, \bibinfo{person}{Li Deng},
  \bibinfo{person}{Jianfeng Gao}, \bibinfo{person}{Xiaodong Liu},
  \bibinfo{person}{Rangan Majumder}, \bibinfo{person}{Andrew McNamara},
  \bibinfo{person}{Bhaskar Mitra}, \bibinfo{person}{Tri Nguyen},
  \bibinfo{person}{Mir Rosenberg}, \bibinfo{person}{Xia Song},
  \bibinfo{person}{Alina Stoica}, \bibinfo{person}{Saurabh Tiwary}, {and}
  \bibinfo{person}{Tong Wang}.} \bibinfo{year}{2016}\natexlab{}.
\newblock \showarticletitle{{MS MARCO}: A Human Generated Machine Reading
  Comprehension Dataset}.
\newblock \bibinfo{journal}{\emph{arXiv preprint arXiv:1611.09268}}
  (\bibinfo{year}{2016}).
\newblock
\showeprint[arxiv]{1611.09268}~[cs.CL]


\bibitem[Burges et~al\mbox{.}(2005)]%
        {Burges05}
\bibfield{author}{\bibinfo{person}{Chris Burges}, \bibinfo{person}{Tal Shaked},
  \bibinfo{person}{Erin Renshaw}, \bibinfo{person}{Ari Lazier},
  \bibinfo{person}{Matt Deeds}, \bibinfo{person}{Nicole Hamilton}, {and}
  \bibinfo{person}{Greg Hullender}.} \bibinfo{year}{2005}\natexlab{}.
\newblock \showarticletitle{Learning to Rank Using Gradient Descent}. In
  \bibinfo{booktitle}{\emph{Proceedings of the 22nd International Conference on
  Machine Learning}}. \bibinfo{publisher}{{ACM}}, \bibinfo{pages}{89--96}.
\newblock


\bibitem[Chen et~al\mbox{.}(2024)]%
        {Chen24BGEM3}
\bibfield{author}{\bibinfo{person}{Jianlv Chen}, \bibinfo{person}{Shitao Xiao},
  \bibinfo{person}{Peitian Zhang}, \bibinfo{person}{Kun Luo},
  \bibinfo{person}{Defu Lian}, {and} \bibinfo{person}{Zheng Liu}.}
  \bibinfo{year}{2024}\natexlab{}.
\newblock \showarticletitle{{BGE M3}-Embedding: Multi-Lingual,
  Multi-Functionality, Multi-Granularity Text Embeddings Through Self-Knowledge
  Distillation}.
\newblock \bibinfo{journal}{\emph{arXiv preprint arXiv:2402.03216}}
  (\bibinfo{year}{2024}).
\newblock
\showeprint[arxiv]{2402.03216}~[cs.CL]


\bibitem[Chen et~al\mbox{.}(2019)]%
        {Chen19}
\bibfield{author}{\bibinfo{person}{Minmin Chen}, \bibinfo{person}{Alex Beutel},
  \bibinfo{person}{Paul Covington}, \bibinfo{person}{Sagar Jain},
  \bibinfo{person}{Francois Belletti}, {and} \bibinfo{person}{Ed~H. Chi}.}
  \bibinfo{year}{2019}\natexlab{}.
\newblock \showarticletitle{Top-{K} Off-Policy Correction for a {REINFORCE}
  Recommender System}. In \bibinfo{booktitle}{\emph{Proceedings of the 12th
  {ACM} International Conference on Web Search and Data Mining}}.
  \bibinfo{publisher}{{ACM}}, \bibinfo{pages}{456--464}.
\newblock
\href{https://doi.org/10.1145/3289600.3290994}{doi:\nolinkurl{10.1145/3289600.3290994}}


\bibitem[Chen and He(2021)]%
        {Chen21}
\bibfield{author}{\bibinfo{person}{Xinlei Chen} {and} \bibinfo{person}{Kaiming
  He}.} \bibinfo{year}{2021}\natexlab{}.
\newblock \showarticletitle{Exploring simple siamese representation learning}.
  In \bibinfo{booktitle}{\emph{Proceedings of the {IEEE/CVF} Conference on
  Computer Vision and Pattern Recognition}}. \bibinfo{publisher}{{IEEE}},
  \bibinfo{pages}{15750--15758}.
\newblock


\bibitem[Covington et~al\mbox{.}(2016)]%
        {Covington16}
\bibfield{author}{\bibinfo{person}{Paul Covington}, \bibinfo{person}{Jay
  Adams}, {and} \bibinfo{person}{Emre Sargin}.}
  \bibinfo{year}{2016}\natexlab{}.
\newblock \showarticletitle{Deep Neural Networks for {YouTube}
  Recommendations}. In \bibinfo{booktitle}{\emph{Proceedings of the 10th {ACM}
  Conference on Recommender Systems}}. \bibinfo{publisher}{{ACM}},
  \bibinfo{pages}{191--198}.
\newblock
\href{https://doi.org/10.1145/2959100.2959190}{doi:\nolinkurl{10.1145/2959100.2959190}}


\bibitem[He et~al\mbox{.}(2020)]%
        {He20}
\bibfield{author}{\bibinfo{person}{Kaiming He}, \bibinfo{person}{Haoqi Fan},
  \bibinfo{person}{Yuxin Wu}, \bibinfo{person}{Saining Xie}, {and}
  \bibinfo{person}{Ross Girshick}.} \bibinfo{year}{2020}\natexlab{}.
\newblock \showarticletitle{Momentum contrast for unsupervised visual
  representation learning}. In \bibinfo{booktitle}{\emph{Proceedings of the
  {IEEE/CVF} Conference on Computer Vision and Pattern Recognition}}.
  \bibinfo{publisher}{{IEEE}}, \bibinfo{pages}{9729--9738}.
\newblock


\bibitem[Huang et~al\mbox{.}(2013)]%
        {Huang13DSSM}
\bibfield{author}{\bibinfo{person}{Po-Sen Huang}, \bibinfo{person}{Xiaodong
  He}, \bibinfo{person}{Jianfeng Gao}, \bibinfo{person}{Li Deng},
  \bibinfo{person}{Alex Acero}, {and} \bibinfo{person}{Larry~P. Heck}.}
  \bibinfo{year}{2013}\natexlab{}.
\newblock \showarticletitle{Learning Deep Structured Semantic Models for Web
  Search Using Clickthrough Data}. In \bibinfo{booktitle}{\emph{Proceedings of
  the 22nd {ACM} International Conference on Information and Knowledge
  Management}}. \bibinfo{publisher}{{ACM}}, \bibinfo{pages}{2333--2338}.
\newblock
\href{https://doi.org/10.1145/2505515.2505665}{doi:\nolinkurl{10.1145/2505515.2505665}}


\bibitem[Johnson et~al\mbox{.}(2021)]%
        {Johnson19}
\bibfield{author}{\bibinfo{person}{Jeff Johnson}, \bibinfo{person}{Matthijs
  Douze}, {and} \bibinfo{person}{Hervé Jégou}.}
  \bibinfo{year}{2021}\natexlab{}.
\newblock \showarticletitle{Billion-scale similarity search with {GPUs}}.
\newblock \bibinfo{journal}{\emph{IEEE Transactions on Big Data}}
  \bibinfo{volume}{7}, \bibinfo{number}{3} (\bibinfo{year}{2021}),
  \bibinfo{pages}{535--547}.
\newblock
\href{https://doi.org/10.1109/TBDATA.2019.2921572}{doi:\nolinkurl{10.1109/TBDATA.2019.2921572}}


\bibitem[Karpukhin et~al\mbox{.}(2020)]%
        {Karpukhin20}
\bibfield{author}{\bibinfo{person}{Vladimir Karpukhin}, \bibinfo{person}{Barlas
  Oguz}, \bibinfo{person}{Sewon Min}, \bibinfo{person}{Patrick Lewis},
  \bibinfo{person}{Ledell Wu}, \bibinfo{person}{Sergey Edunov},
  \bibinfo{person}{Danqi Chen}, {and} \bibinfo{person}{Wen tau Yih}.}
  \bibinfo{year}{2020}\natexlab{}.
\newblock \showarticletitle{Dense Passage Retrieval for Open-Domain Question
  Answering}. In \bibinfo{booktitle}{\emph{Proceedings of the 2020 Conference
  on Empirical Methods in Natural Language Processing}}.
  \bibinfo{publisher}{Association for Computational Linguistics},
  \bibinfo{pages}{6769--6781}.
\newblock


\bibitem[Li et~al\mbox{.}(2023)]%
        {Li23GTE}
\bibfield{author}{\bibinfo{person}{Zehan Li}, \bibinfo{person}{Xin Zhang},
  \bibinfo{person}{Yanzhao Zhang}, \bibinfo{person}{Dingkun Long},
  \bibinfo{person}{Pengjun Xie}, {and} \bibinfo{person}{Meishan Zhang}.}
  \bibinfo{year}{2023}\natexlab{}.
\newblock \showarticletitle{Towards General Text Embeddings with Multi-stage
  Contrastive Learning}.
\newblock \bibinfo{journal}{\emph{arXiv preprint arXiv:2308.03281}}
  (\bibinfo{year}{2023}).
\newblock
\showeprint[arxiv]{2308.03281}~[cs.CL]


\bibitem[Lin et~al\mbox{.}(2021)]%
        {Lin21}
\bibfield{author}{\bibinfo{person}{Jimmy Lin}, \bibinfo{person}{Rodrigo
  Nogueira}, {and} \bibinfo{person}{Andrew Yates}.}
  \bibinfo{year}{2021}\natexlab{}.
\newblock \bibinfo{booktitle}{\emph{Pretrained Transformers for Text Ranking:
  {BERT} and Beyond}}.
\newblock \bibinfo{publisher}{Morgan \& Claypool}.
\newblock


\bibitem[Lin et~al\mbox{.}(2025)]%
        {Lin25}
\bibfield{author}{\bibinfo{person}{Zhijie Lin}, \bibinfo{person}{Zhuofeng Li},
  \bibinfo{person}{Chenglei Dai}, \bibinfo{person}{Wentian Bao},
  \bibinfo{person}{Shuai Lin}, \bibinfo{person}{Enyun Yu},
  \bibinfo{person}{Haoxiang Zhang}, {and} \bibinfo{person}{Liang Zhao}.}
  \bibinfo{year}{2025}\natexlab{}.
\newblock \showarticletitle{{GReF}: A Unified Generative Framework for
  Efficient Reranking via Ordered Multi-token Prediction}. In
  \bibinfo{booktitle}{\emph{Proceedings of the 34th {ACM} International
  Conference on Information and Knowledge Management}}.
  \bibinfo{publisher}{{ACM}}, \bibinfo{pages}{5879--5887}.
\newblock
\href{https://doi.org/10.1145/3746252.3761540}{doi:\nolinkurl{10.1145/3746252.3761540}}


\bibitem[Liu et~al\mbox{.}(2025)]%
        {Tao25}
\bibfield{author}{\bibinfo{person}{Xingxian Liu}, \bibinfo{person}{Dongshuai
  Li}, \bibinfo{person}{Jiahui Wan}, \bibinfo{person}{Tao Wen},
  \bibinfo{person}{Gui Ling}, \bibinfo{person}{Yuliang Yan},
  \bibinfo{person}{Fuyu Lv}, \bibinfo{person}{Dan Ou}, \bibinfo{person}{Haihong
  Tang}, {and} \bibinfo{person}{Bo Zheng}.} \bibinfo{year}{2025}\natexlab{}.
\newblock \showarticletitle{{Retrieval-GRPO}: A Multi-Objective Reinforcement
  Learning Framework for Dense Retrieval in Taobao Search}.
\newblock \bibinfo{journal}{\emph{arXiv preprint arXiv:2511.13885}}
  (\bibinfo{year}{2025}).
\newblock
\showeprint[arxiv]{2511.13885}~[cs.IR]


\bibitem[Nogueira and Cho(2019)]%
        {Nogueira19}
\bibfield{author}{\bibinfo{person}{Rodrigo Nogueira} {and}
  \bibinfo{person}{Kyunghyun Cho}.} \bibinfo{year}{2019}\natexlab{}.
\newblock \showarticletitle{Passage Re-ranking with {BERT}}.
\newblock \bibinfo{journal}{\emph{arXiv preprint arXiv:1901.04085}}
  (\bibinfo{year}{2019}).
\newblock
\showeprint[arxiv]{1901.04085}~[cs.IR]


\bibitem[Oosterhuis and de~Rijke(2018)]%
        {Oosterhuis18}
\bibfield{author}{\bibinfo{person}{Harrie Oosterhuis} {and}
  \bibinfo{person}{Maarten de Rijke}.} \bibinfo{year}{2018}\natexlab{}.
\newblock \showarticletitle{Differentiable unbiased online learning to rank}.
  In \bibinfo{booktitle}{\emph{Proceedings of the 27th {ACM} International
  Conference on Information and Knowledge Management}}.
  \bibinfo{publisher}{{ACM}}, \bibinfo{pages}{1293--1302}.
\newblock


\bibitem[Ouyang et~al\mbox{.}(2022)]%
        {Ouyang22}
\bibfield{author}{\bibinfo{person}{Long Ouyang}, \bibinfo{person}{Jeff Wu},
  \bibinfo{person}{Xu Jiang}, \bibinfo{person}{Diogo Almeida},
  \bibinfo{person}{Carroll~L. Wainwright}, \bibinfo{person}{Pamela Mishkin},
  \bibinfo{person}{Chong Zhang}, \bibinfo{person}{Sandhini Agarwal},
  \bibinfo{person}{Katarina Slama}, \bibinfo{person}{Alex Ray},
  \bibinfo{person}{John Schulman}, \bibinfo{person}{Jacob Hilton},
  \bibinfo{person}{Fraser Kelton}, \bibinfo{person}{Luke Miller},
  \bibinfo{person}{Maddie Simens}, \bibinfo{person}{Amanda Askell},
  \bibinfo{person}{Peter Welinder}, \bibinfo{person}{Paul Christiano},
  \bibinfo{person}{Jan Leike}, {and} \bibinfo{person}{Ryan Lowe}.}
  \bibinfo{year}{2022}\natexlab{}.
\newblock \showarticletitle{Training Language Models to Follow Instructions
  with Human Feedback}.
\newblock \bibinfo{journal}{\emph{Advances in Neural Information Processing
  Systems}}  \bibinfo{volume}{35} (\bibinfo{year}{2022}),
  \bibinfo{pages}{27730--27744}.
\newblock


\bibitem[Qu et~al\mbox{.}(2021)]%
        {Qu21}
\bibfield{author}{\bibinfo{person}{Yingqi Qu}, \bibinfo{person}{Yuchen Ding},
  \bibinfo{person}{Jing Liu}, \bibinfo{person}{Kai Liu},
  \bibinfo{person}{Ruiyang Ren}, \bibinfo{person}{Wayne~Xin Zhao},
  \bibinfo{person}{Daxiang Dong}, \bibinfo{person}{Hua Wu}, {and}
  \bibinfo{person}{Haifeng Wang}.} \bibinfo{year}{2021}\natexlab{}.
\newblock \showarticletitle{{RocketQA}: An Optimized Training Approach to Dense
  Passage Retrieval for Open-Domain Question Answering}. In
  \bibinfo{booktitle}{\emph{Proceedings of the 2021 Conference of the North
  American Chapter of the Association for Computational Linguistics: Human
  Language Technologies}}. \bibinfo{publisher}{Association for Computational
  Linguistics}, \bibinfo{pages}{5835--5847}.
\newblock


\bibitem[{Qwen Team}(2025)]%
        {Yang25}
\bibfield{author}{\bibinfo{person}{{Qwen Team}}.}
  \bibinfo{year}{2025}\natexlab{}.
\newblock \showarticletitle{{Qwen3} Technical Report}.
\newblock \bibinfo{journal}{\emph{arXiv preprint arXiv:2505.09388}}
  (\bibinfo{year}{2025}).
\newblock
\showeprint[arxiv]{2505.09388}~[cs.CL]
\href{https://doi.org/10.48550/arXiv.2505.09388}{doi:\nolinkurl{10.48550/arXiv.2505.09388}}


\bibitem[Ren et~al\mbox{.}(2021)]%
        {Ren21}
\bibfield{author}{\bibinfo{person}{Ruiyang Ren}, \bibinfo{person}{Yingqi Qu},
  \bibinfo{person}{Jing Liu}, \bibinfo{person}{Wayne~Xin Zhao},
  \bibinfo{person}{Qiaoqiao She}, \bibinfo{person}{Hua Wu},
  \bibinfo{person}{Haifeng Wang}, {and} \bibinfo{person}{Ji-Rong Wen}.}
  \bibinfo{year}{2021}\natexlab{}.
\newblock \showarticletitle{{RocketQAv2}: A Joint Training Method for Dense
  Passage Retrieval and Passage Re-ranking}. In
  \bibinfo{booktitle}{\emph{Proceedings of the 2021 Conference on Empirical
  Methods in Natural Language Processing}}. \bibinfo{publisher}{Association for
  Computational Linguistics}, \bibinfo{pages}{2825--2835}.
\newblock


\bibitem[Shao et~al\mbox{.}(2024)]%
        {Shao24}
\bibfield{author}{\bibinfo{person}{Zhihong Shao}, \bibinfo{person}{Peiyi Wang},
  \bibinfo{person}{Qihao Zhu}, \bibinfo{person}{Runxin Xu},
  \bibinfo{person}{Junxiao Song}, \bibinfo{person}{Xiao Bi},
  \bibinfo{person}{Haowei Zhang}, \bibinfo{person}{Mingchuan Zhang},
  \bibinfo{person}{Y.~K. Li}, \bibinfo{person}{Y. Wu}, {and}
  \bibinfo{person}{Daya Guo}.} \bibinfo{year}{2024}\natexlab{}.
\newblock \showarticletitle{{DeepSeekMath}: Pushing the Limits of Mathematical
  Reasoning in Open Language Models}.
\newblock \bibinfo{journal}{\emph{arXiv preprint arXiv:2402.03300}}
  (\bibinfo{year}{2024}).
\newblock
\showeprint[arxiv]{2402.03300}~[cs.CL]


\bibitem[Singh and Joachims(2018)]%
        {Singh18}
\bibfield{author}{\bibinfo{person}{A. Singh} {and} \bibinfo{person}{T.
  Joachims}.} \bibinfo{year}{2018}\natexlab{}.
\newblock \showarticletitle{Fairness of Exposure in Rankings}. In
  \bibinfo{booktitle}{\emph{Proceedings of the 24th {ACM} {SIGKDD}
  International Conference on Knowledge Discovery {\&} Data Mining}}.
  \bibinfo{publisher}{{ACM}}, \bibinfo{pages}{2219--2228}.
\newblock
\href{https://doi.org/10.1145/3219819.3220078}{doi:\nolinkurl{10.1145/3219819.3220078}}


\bibitem[Sutton and Barto(2018)]%
        {Sutton18}
\bibfield{author}{\bibinfo{person}{Richard~S. Sutton} {and}
  \bibinfo{person}{Andrew~G. Barto}.} \bibinfo{year}{2018}\natexlab{}.
\newblock \bibinfo{booktitle}{\emph{Reinforcement Learning: An Introduction}
  (\bibinfo{edition}{2nd} ed.)}.
\newblock \bibinfo{publisher}{{MIT} Press}.
\newblock
\showISBNx{978-0262039246}


\bibitem[van~den Oord et~al\mbox{.}(2018)]%
        {Oord18}
\bibfield{author}{\bibinfo{person}{Aaron van~den Oord}, \bibinfo{person}{Yazhe
  Li}, {and} \bibinfo{person}{Oriol Vinyals}.} \bibinfo{year}{2018}\natexlab{}.
\newblock \showarticletitle{Representation Learning with Contrastive Predictive
  Coding}.
\newblock \bibinfo{journal}{\emph{arXiv preprint arXiv:1807.03748}}
  (\bibinfo{year}{2018}).
\newblock
\showeprint[arxiv]{1807.03748}~[cs.LG]


\bibitem[van~der Maaten and Hinton(2008)]%
        {Maaten08}
\bibfield{author}{\bibinfo{person}{Laurens van~der Maaten} {and}
  \bibinfo{person}{Geoffrey Hinton}.} \bibinfo{year}{2008}\natexlab{}.
\newblock \showarticletitle{Visualizing Data Using {t-SNE}}.
\newblock \bibinfo{journal}{\emph{Journal of Machine Learning Research}}
  \bibinfo{volume}{9} (\bibinfo{date}{Nov} \bibinfo{year}{2008}),
  \bibinfo{pages}{2579--2605}.
\newblock


\bibitem[Williams(1992)]%
        {Williams92}
\bibfield{author}{\bibinfo{person}{Ronald~J. Williams}.}
  \bibinfo{year}{1992}\natexlab{}.
\newblock \showarticletitle{Simple statistical gradient-following algorithms
  for connectionist reinforcement learning}.
\newblock \bibinfo{journal}{\emph{Machine Learning}} \bibinfo{volume}{8},
  \bibinfo{number}{3} (\bibinfo{year}{1992}), \bibinfo{pages}{229--256}.
\newblock


\bibitem[Xiao et~al\mbox{.}(2023)]%
        {Xiao23}
\bibfield{author}{\bibinfo{person}{Shitao Xiao}, \bibinfo{person}{Zheng Liu},
  \bibinfo{person}{Peitian Zhang}, \bibinfo{person}{Niklas Muennighoff},
  \bibinfo{person}{Defu Lian}, {and} \bibinfo{person}{Jian-Yun Nie}.}
  \bibinfo{year}{2023}\natexlab{}.
\newblock \showarticletitle{{C-Pack}: Packed Resources For General Chinese
  Embeddings}.
\newblock \bibinfo{journal}{\emph{arXiv preprint arXiv:2309.07597}}
  (\bibinfo{year}{2023}).
\newblock
\showeprint[arxiv]{2309.07597}~[cs.CL]


\bibitem[Xiong et~al\mbox{.}(2021)]%
        {Xiong20}
\bibfield{author}{\bibinfo{person}{Lee Xiong}, \bibinfo{person}{Chenyan Xiong},
  \bibinfo{person}{Ye Li}, \bibinfo{person}{Kwok-Fung Tang},
  \bibinfo{person}{Jialin Liu}, \bibinfo{person}{Paul Bennett},
  \bibinfo{person}{Junaid Ahmed}, {and} \bibinfo{person}{Arnold Overwijk}.}
  \bibinfo{year}{2021}\natexlab{}.
\newblock \showarticletitle{Approximate Nearest Neighbor Negative Contrastive
  Learning for Dense Text Retrieval}. In
  \bibinfo{booktitle}{\emph{International Conference on Learning
  Representations}}.
\newblock
\showeprint[arxiv]{2007.00808}~[cs.IR]


\bibitem[Xu et~al\mbox{.}(2025)]%
        {Xu25}
\bibfield{author}{\bibinfo{person}{Bo Xu}, \bibinfo{person}{Yicen Tian},
  \bibinfo{person}{Xiaokun Zhang}, \bibinfo{person}{Erchen Yu},
  \bibinfo{person}{Dailin Li}, \bibinfo{person}{Linlin Zong}, {and}
  \bibinfo{person}{Hongfei Lin}.} \bibinfo{year}{2025}\natexlab{}.
\newblock \showarticletitle{Reinforcement Learning-Driven Generative Retrieval
  with Semantic-aligned Multi-Layer Identifiers}. In
  \bibinfo{booktitle}{\emph{Proceedings of the 34th {ACM} International
  Conference on Information and Knowledge Management}}.
  \bibinfo{publisher}{{ACM}}, \bibinfo{pages}{3592--3601}.
\newblock
\href{https://doi.org/10.1145/3746252.3761136}{doi:\nolinkurl{10.1145/3746252.3761136}}


\bibitem[Zeng et~al\mbox{.}(2020)]%
        {Zeng20}
\bibfield{author}{\bibinfo{person}{Hansi Zeng}, \bibinfo{person}{Hamed Zamani},
  {and} \bibinfo{person}{W.~Bruce Croft}.} \bibinfo{year}{2020}\natexlab{}.
\newblock \showarticletitle{Curriculum Learning for Dense Retrieval
  Distillation}. In \bibinfo{booktitle}{\emph{Proceedings of the 43rd
  International {ACM} {SIGIR} Conference on Research and Development in
  Information Retrieval}}. \bibinfo{publisher}{{ACM}},
  \bibinfo{pages}{1979--1982}.
\newblock


\bibitem[Zhang et~al\mbox{.}(2024)]%
        {Zhang24}
\bibfield{author}{\bibinfo{person}{Xin Zhang}, \bibinfo{person}{Yanzhao Zhang},
  \bibinfo{person}{Dingkun Long}, \bibinfo{person}{Wen Xie},
  \bibinfo{person}{Ziqi Dai}, \bibinfo{person}{Jialong Tang},
  \bibinfo{person}{Huan Lin}, \bibinfo{person}{Baosong Yang},
  \bibinfo{person}{Pengjun Xie}, \bibinfo{person}{Fei Huang},
  \bibinfo{person}{Meishan Zhang}, \bibinfo{person}{Wenjie Li}, {and}
  \bibinfo{person}{Min Zhang}.} \bibinfo{year}{2024}\natexlab{}.
\newblock \showarticletitle{{mGTE}: Generalized Long-Context Text
  Representation and Reranking Models for Multilingual Text Retrieval}.
\newblock \bibinfo{journal}{\emph{arXiv preprint arXiv:2407.19669}}
  (\bibinfo{year}{2024}).
\newblock
\showeprint[arxiv]{2407.19669}~[cs.CL]


\bibitem[Zhuang et~al\mbox{.}(2026)]%
        {Zhuang26RankR1}
\bibfield{author}{\bibinfo{person}{Shengyao Zhuang}, \bibinfo{person}{Xueguang
  Ma}, \bibinfo{person}{Zheng Yao}, \bibinfo{person}{Shuai Wang},
  \bibinfo{person}{Bevan Koopman}, \bibinfo{person}{Jimmy Lin}, {and}
  \bibinfo{person}{Guido Zuccon}.} \bibinfo{year}{2026}\natexlab{}.
\newblock \showarticletitle{Rank-{R1}: Enhancing Reasoning in {LLM}-based
  Document Rerankers via Reinforcement Learning}. In
  \bibinfo{booktitle}{\emph{Proceedings of the 49th International {ACM} {SIGIR}
  Conference on Research and Development in Information Retrieval}}.
  \bibinfo{publisher}{{ACM}}, \bibinfo{pages}{1--7}.
\newblock
\href{https://doi.org/10.1145/3805712.3809961}{doi:\nolinkurl{10.1145/3805712.3809961}}


\end{thebibliography}

\end{document}